\documentclass[preprint,12pt]{elsarticle}

\usepackage{amssymb,amsfonts,amsmath}
\usepackage{graphicx}
\usepackage{dcolumn}
\usepackage{color}

\journal{Physica A}

\begin{document}

\begin{frontmatter}

\title{Can we neglect the multi-layer structure of functional networks?}

\author{Massimiliano Zanin}

\address{Innaxis Foundation \& Research Institute,
Jos\'e Ortega y Gasset 20, 28006, Madrid, Spain}
\address{Faculdade de Ci\^encias e Tecnologia, Departamento de Engenharia Electrot\'ecnica,
Universidade Nova de Lisboa, Lisboa, Portugal}

\begin{abstract}
Functional networks, {\it i.e.} networks representing dynamic relationships between the components of a complex system, have been instrumental for our understanding of, among others, the human brain. Due to limited data availability, the multi-layer nature of numerous functional networks has hitherto been neglected, and nodes are endowed with a single type of links even when multiple relationships coexist at different physical levels. A relevant problem is the assessment of the benefits yielded by studying a multi-layer functional network, against the simplicity guaranteed by the reconstruction and use of the corresponding single layer projection. Here, I tackle this issue by using as a test case, the functional network representing the dynamics of delay propagation through European airports. Neglecting the multi-layer structure of a functional network has dramatic consequences on our understanding of the underlying system, a fact to be taken into account when a projection is the only available information.
\end{abstract}


\end{frontmatter}

\section{Introduction}

In the early stages of complex network theory \cite{newman2003,boccaletti2006}, such paradigm was mainly used to analyze systems whose structure, either physical or virtual, could be directly mapped into a network: examples include transportation systems \cite{sienkiewicz2005,zanin2013}, social networks \cite{wasserman1994} or the World Wide Web \cite{barabasi1999}. 
It was soon clear that in certain cases this was not possible, as the only information obtainable from the system itself was the evolution through time of some observables. Such measurable variables reflect the behavior of the interacting elements constituting the system, and as such, the value of every observable is expected to be a {\it function} of the values of other peers. When the structure of such interactions is inferred from the dynamics of the observables, the result is then called a {\it functional network}. 
The introduction of this latter representation has resulted in an important step forward in network science, allowing a broader focus including both structural and dynamical (functional) relations.

Among the examples of functional network representations, the study of the brain \cite{bullmore2009,rubinov2010} and of gene expressions \cite{stuart2003, oldham2006} are probably the most paradigmatic. In the former, nodes correspond to sensors of a machine recording the activity of the brain. Through the magnetic or electric field generated by spiking neurons, links are then established whenever some kind of synchronization is detected between the recorded time series, usually by means of metrics like Pearson's linear correlation, Synchronization Likelihood \cite{stam2002}, or Granger Causality \cite{granger1969}. Similarly, nodes in gene co-expression networks represent individual genes, pairwise connected when some correlation is detected in the dynamics of their respective expression levels.

In the last few years researchers have realized that interactions between the constituting elements of complex systems seldom develop on a single channel. 
For instance, in a social network, information exchange may happen orally, electronically, or even indirectly. Additionally, people interact according to different types of relationships, {\it e.g.}~friendship and co-working, each one of these affecting the type of information transmitted. Consequently, a correct representation may require different types, or layers, of links \cite{boccaletti2014structure}. Neglecting such multi-layer structure, or in other words working with the {\it projected network}, may alter our perception of the topology and dynamics, leading to a wrong understanding of the properties of the system \cite{buldyrev2010,vespignani2010}. In such cases, therefore, a single-layer network may be an oversimplification, and a multi-layer structure is required.

A single-layer functional representation is generally created by collapsing the dynamics of the individual elements, such that the multiple dynamical aspects of each node are merged into a single time evolution. The same applies to the case of functional brain networks: the six-layer structure of the human cortex is neglected due to the limited spatial resolution of magnetic and electric sensors, and the analyzed time series correspond to the global activity of the top-most layers.
Nevertheless, the non-linear nature of the projection process can foster the appearance of constructive or destructive interferences. In other words, a link may appear in the projection even if no relationship is present in any layer; or links in two layers can interfere, and disappear from the projection.

If one is to represent a real system by means of a functional network, and he / she expects it to have a multi-layer structure, it would then be important to understand the magnitude of the distortions created by the use of a single-layer representation.
In other words, a fundamental question should be addressed: to what degree single-layer functional networks are representative of the dynamics occurring at different layers?

In order to address this question, here I analyze the European air transportation network and create functional networks modeling the influence airports have on each another, in terms of flight delays and propagation.
The availability of high-resolution real data allows the reconstruction of a complete multi-layer picture, in which each layer corresponds to a different airline; dynamics are afterwards collapsed, in order to simulate the creation of a single-layer representation. The resulting structures are compared, both topologically and dynamically, in order to assess whether they provide similar insight into the underlying system.

\section{Reconstructing the air transport functional network}

The air transportation system can be described both from a physical and a functional perspective. According to the former, the system supports a transport phenomenon, in which passengers and goods are transported between pairs of airports. Therefore, the corresponding network representation encodes the presence of direct connections between nodes, in this case direct flights between airports. Numerous works have focused on this aspect, as for instance Refs. \cite{guimera2005,colizza2006,zanin2013}. 
A latter perspective disregards the physical movement of items in the network, focusing instead on how the dynamics of nodes is influenced by other peers. Thus, a time evolving observable characterizes each element of the system, its dynamics being a function of the dynamics of its neighboring elements. 
When such functional relationships are uncovered, the result is a {\it functional} network. The air transport system can be seen as a collection of airports, whose observables are the delay experienced in each one of them. Clearly, delays are transported by aircraft, and thus the delay observed at one airport is a (potentially noisy) function of other airports delays. Such relationships are then susceptible of being mapped in a functional complex network. 

The reconstruction of the physical network is a straightforward process, only requiring an analysis of flights' schedules.
On the other hand, the reconstruction of delay functional networks starts with time series representing average landing delays across European airports. 
In this example, I have considered delay time series for the $50$ busiest European airports and the $20$ largest airlines in number of flights during $10$ consecutive months of year $2011$ - see \ref{app:data} for further details. These time series have been pre-processed, in order to ensure their stationarity, for then assessing the synchronization level between pairs of them (see \ref{app:TSExtraction}).

\begin{figure}[!tb]
\begin{center}
\includegraphics[width=0.99\textwidth]{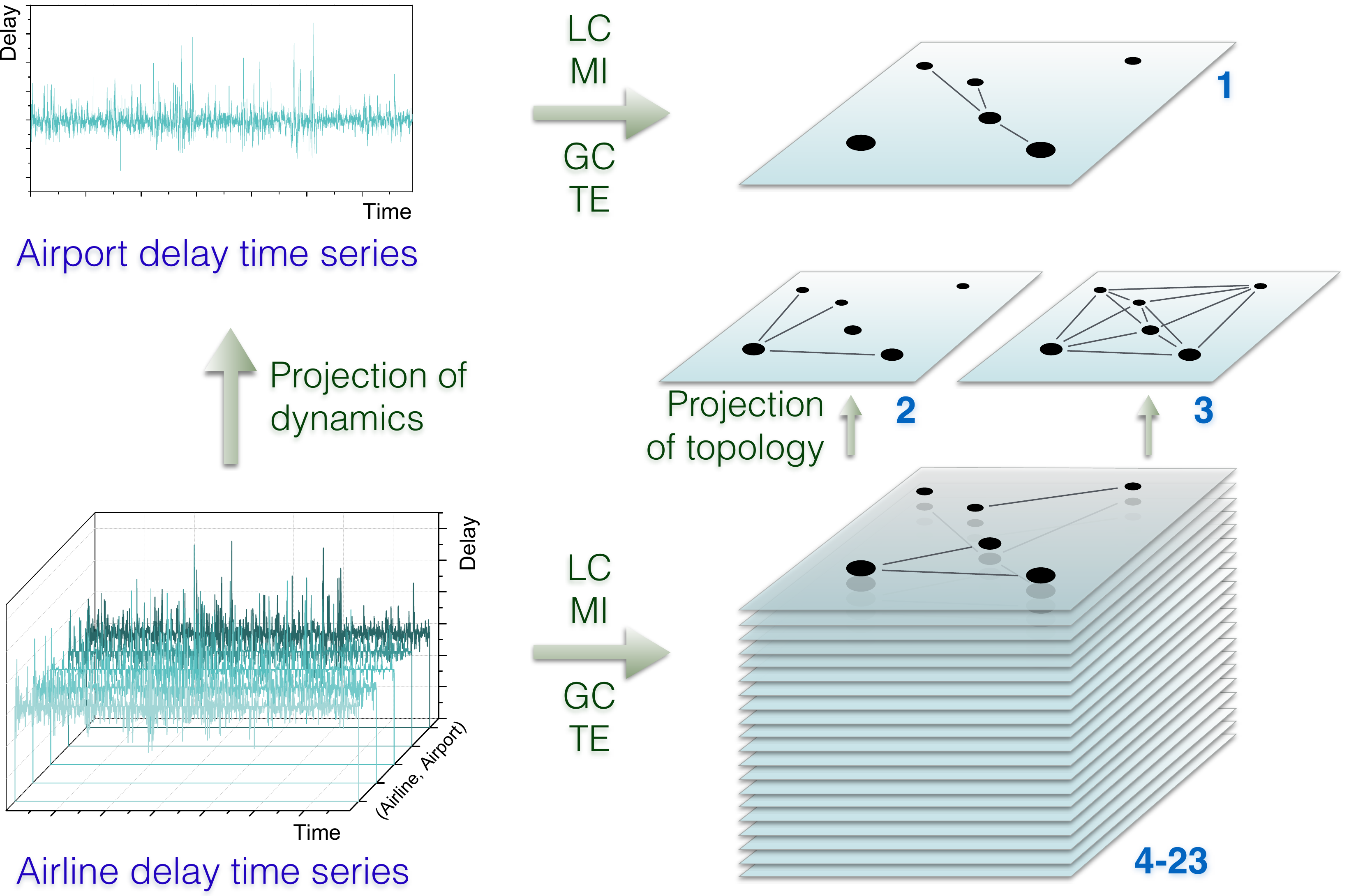}
\caption{Representation of the network reconstruction procedure. Starting from raw delay time series corresponding to different airports and airlines (bottom left), a multi-layer representation is created (bottom right). Furthermore, two projections are considered: a projection of the dynamics, where delays of airports are collapsed to create a single layer functional network (top left, and network number $1$); and a topological projection of the multi-layer representation (networks number $2$ and $3$). In network $3$, all links are considered, while in network $2$ links are pruned to obtain the same link density as in network $1$. {\it LC}, {\it MI}, {\it GC} and {\it TE} stand for the four synchronization metrics considered, as described in \ref{app:TSExtraction}.}\label{fig:01}
\end{center}
\end{figure}

As depicted in Fig. \ref{fig:01}, by starting from the raw time series (bottom left part), two possibilities can be followed. On one side, one can average all of the dynamics corresponding to a node, or in this case to an airport, thus creating a network representing the {\it projection of the dynamics}. Once one time series per airport is available, different metrics can then be calculated to construct a weighted fully-connected network. On the other side, one can directly create a multi-layer network by considering each airline as a layer, in which afterwards such structure can be projected into a single-layer graph, thus creating a {\it projection of the topology}.

According to this idea, $23$ different networks can be created: following the numeration of Fig. \ref{fig:01}, a {\it projection of the dynamics} ($1$); two {\it projections of the topology}, by respectively taking into account the strongest links in each layer ($2$) and the sum all statistically significant links ($3$), and the $20$ airline networks ($4 - 23$). All networks, except for number $3$, have been pruned in order to have a link density of $0.05$.

\section{Topological properties}

Differences between the $23$ reconstructed networks are first assessed from a topological point of view. Considering that this is a diffusion process involving delays on top of the European air transportation network, the following topological metrics have been selected for their relevance:

\begin{description}
\item[Out-degree centrality,] {\it i.e.} the number of links departing from a given node. This centrality assesses the capacity of a node of inducing delays in neighbor airports.

\item[$\alpha$-centrality,] defining the centrality of a node as the result of a random walk in the graph \cite{bonacich2001eigenvector}.

\item[Efficiency,] defined as the inverse of the harmonic average of the geodesic distance between all pairs of nodes, and measuring how efficiently the network can diffuse information \cite{latora2001}.

\item[Clustering coefficient,] a measure assessing the propensity of nodes to cluster together, and defined as the number of closed triplets ({\it i.e.} the number of triangles in the network) over the total number of connected triplets.

\item[Degree-degree correlation,] measuring the assortativity of the network, {\it i.e.} the fact that nodes tend to connect to their connectivity peers; it is calculated through the Pearson correlation coefficient of the degrees at either ends of a link \cite{newman2002assortative}.

\item[Entropy of the degree distribution,] quantifying the network heterogeneity, and related with its robustness against random failures \cite{wang2006entropy}.

\item[Normalized Information Content (IC),] assessing the presence of regularities in the adjacency matrix, and thus the presence of meso-scale structures \cite{zanin2014information}. The lower the value, the more regular is the link arrangement, indicating the presence of meso-scale structures.

\end{description}

\begin{table*}[!tb]
\begin{center}
  
  \caption{\label{tab:topology} Main topological properties of the reconstructed networks. From left to right, the six columns represent the following networks: the projection of the dynamics (network $1$ of Fig. \ref{fig:01}), the two projections of the topology ($2$ and $3$ of Fig. \ref{fig:01}), and the first three layers (airlines).}
\scriptsize
  \begin{tabular}{ | l | l | l | l | l | l | l | }
    \hline

        & {\bf Proj. (1)}    & {\bf Proj. (2)}    & {\bf Proj. (3)}    &    {\bf Layer 1}    & {\bf Layer 2}    & {\bf Layer 3}   \\ \hline

\rule{0pt}{3ex} 

{\bf Link density}              & 0.05   & 0.05   & 0.54   & 0.05   & 0.05   & 0.05   \\
\rule{0pt}{4ex} 
{\bf Maximum out degree}               &   26   &   16   & 48     & 13     & 13     & 15     \\
\rule{0pt}{4ex} 
{\bf Degree-degree correlation} & -0.058 & -0.077 & -0.029 & 0.031  & -0.088 & 0.094  \\
\rule{0pt}{4ex} 
{\bf Clustering Coefficient}    & 0.022  & 0.179  & 0.707  & 0.064  & 0.217  & 0.095  \\
\rule{0pt}{4ex} 
{\bf Efficiency}                & 0.064  & 0.094  & 0.765  & 0.110  & 0.064  & 0.113  \\
\rule{0pt}{4ex} 
{\bf Normalized IC}             & 0.875  & 0.109  & 0.179  & 0.463  & 0.298  & 0.216  \\
\rule{0pt}{4ex} 
{\bf Size giant component}      & 2      & 17     & 48     & 16     & 14     & 18   \\
    \hline
  \end{tabular}
\end{center}
\end{table*}

\begin{figure*}[!tb]
\begin{center}
\includegraphics[width=0.99\textwidth]{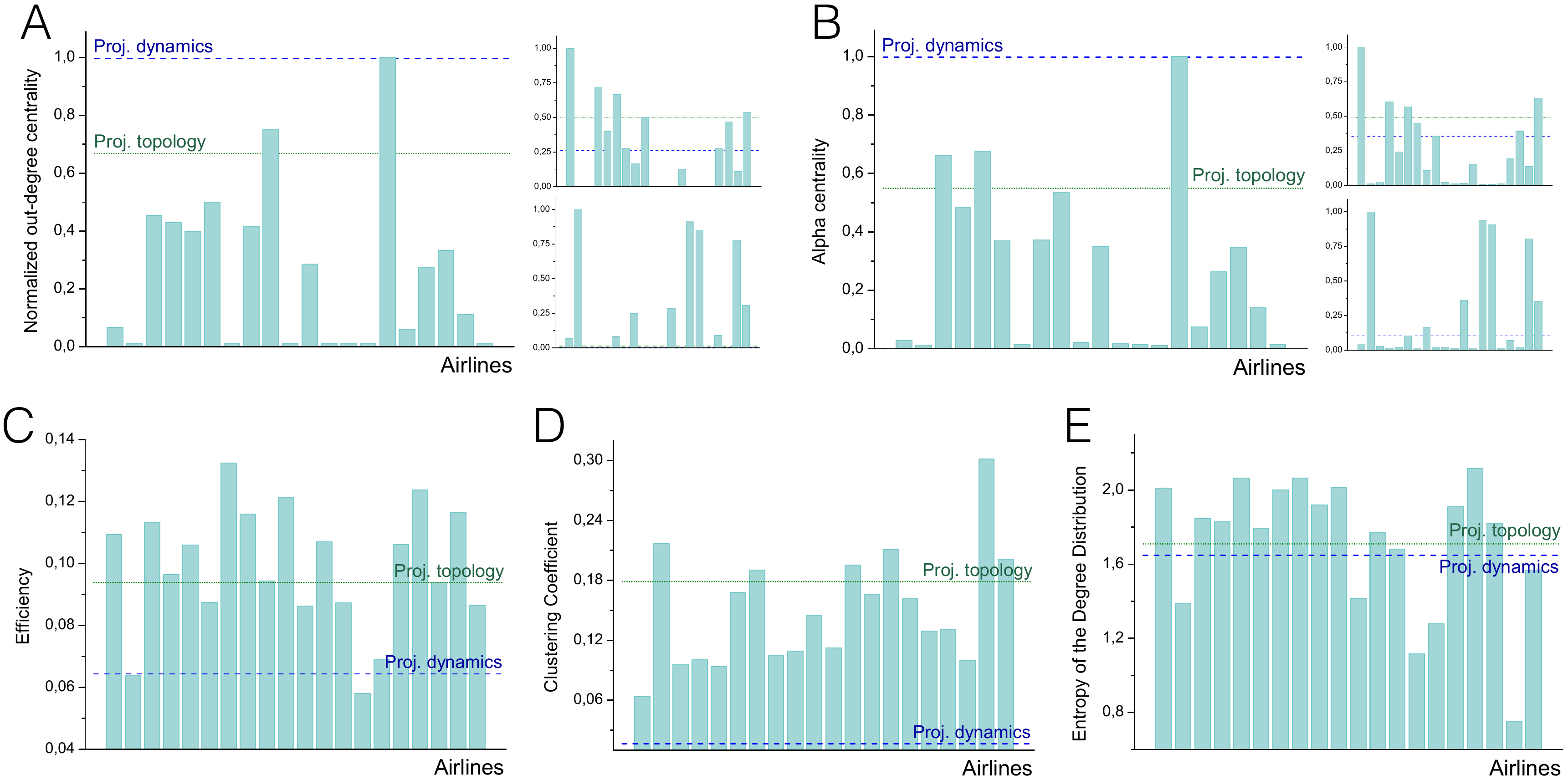}
\caption{Characterization of the functional networks. ({\it A}) and ({\it B}) respectively represent the out-degree centrality and $\alpha$-centrality of the most central node in the dynamical projection (network $1$ of Fig. \ref{fig:01}) in the network corresponding to the projected topology (network $2$ of Fig. \ref{fig:01}) and in all layers of the multi-layer representation. The two small graphs on the right respectively represent the same results for the most central node for the first and second layer. ({\it C}), ({\it D}) and ({\it E}) respectively represent the efficiency, clustering coefficient and entropy of the degree distribution for the two projected functional networks and the layers of the multi-layer representation.}\label{fig:02}
\end{center}
\end{figure*}

Table \ref{tab:topology} and Fig. \ref{fig:02} report the main the results obtained in all considered networks. The two networks respectively obtained by projecting the dynamics and the topology of the system (number $1$ and $2$ of Fig. \ref{fig:01}) share a $90.5\%$ of their links. Nevertheless, they are characterized by different topologies, as shown in Tab. \ref{tab:topology}. Specifically, the former has a centralized structure, as indicated by the higher maximum degree; the latter is less centralized, with a higher clustering coefficient.
Due to the high number of links composing it, the network representing the unpruned projection of the topology (number $3$ of Fig. \ref{fig:01}) is the most different, sharing respectively only a $45.1\%$ and $50.65\%$ of links with the two peers. Such high connectiveness reflects in the topological metrics, yielding high clustering coefficient and efficiency.
Finally, Tab. \ref{tab:topology} reports some topological characteristics of the first three layers, corresponding to the three most important airlines operating in Europe. While the first and third one share most characteristics, the second one is characterized by an higher clustering coefficient; this is due to their different business models, being the first and the third major airlines, the second a low-cost.

Figs. \ref{fig:02} A and B focus on the distribution of node centralities, thus indicating which nodes are most relevant from the point of view of delay propagation. In both panels, the main graph reports the centrality of the most important node in the network resulting from the projection of the dynamics (network number $1$ in Fig. \ref{fig:01}), as calculated by means of the out-degree centrality (panel A) and $\alpha$-centrality (panel B). The two horizontal lines and the vertical bars respectively represent the centrality of that node in the projection of the dynamics (by definition, equal to $1$), in the projection of the topology (network $2$ in Fig. \ref{fig:01}), and in all the $20$ layers of the multi-layer representation. Furthermore, the two small graphs on the right part depict the same information for the most central node in the first two layers. Figs. \ref{fig:02} C, D and E report on three additional topological properties for the $22$ considered networks, {\it i.e.} efficiency, clustering coefficient and entropy of the degree distribution.

In each case, a strong heterogeneity in results is obtained. Specifically, the most important node in one network is seldom of major relevance in other networks. Also, topological features like efficiency and clustering coefficient take significantly different values depending on the type of projection considered.
This has important consequences from an engineering point of view. For instance, if one is to identify the most important node (airport) in terms of delay propagation, the use of different projections would yield different results, providing no guarantee that the most central node in the projected networks is indeed relevant when the multi-layer dynamics is considered.

\section{Dynamical properties}

From a dynamical perspective, it is necessary to ascertain whether such topological differences are enough to alter our understanding of the dynamics of the system. 
Specifically, in the case of air transport, this involves quantifying the amount of delay generated within the system, as well as its spatial and temporal evolution; on the other side, estimating the error resulting from the use of a projection of the multi-layer structure.

\begin{figure*}[!tb]
\begin{center}
\includegraphics[width=0.99\textwidth]{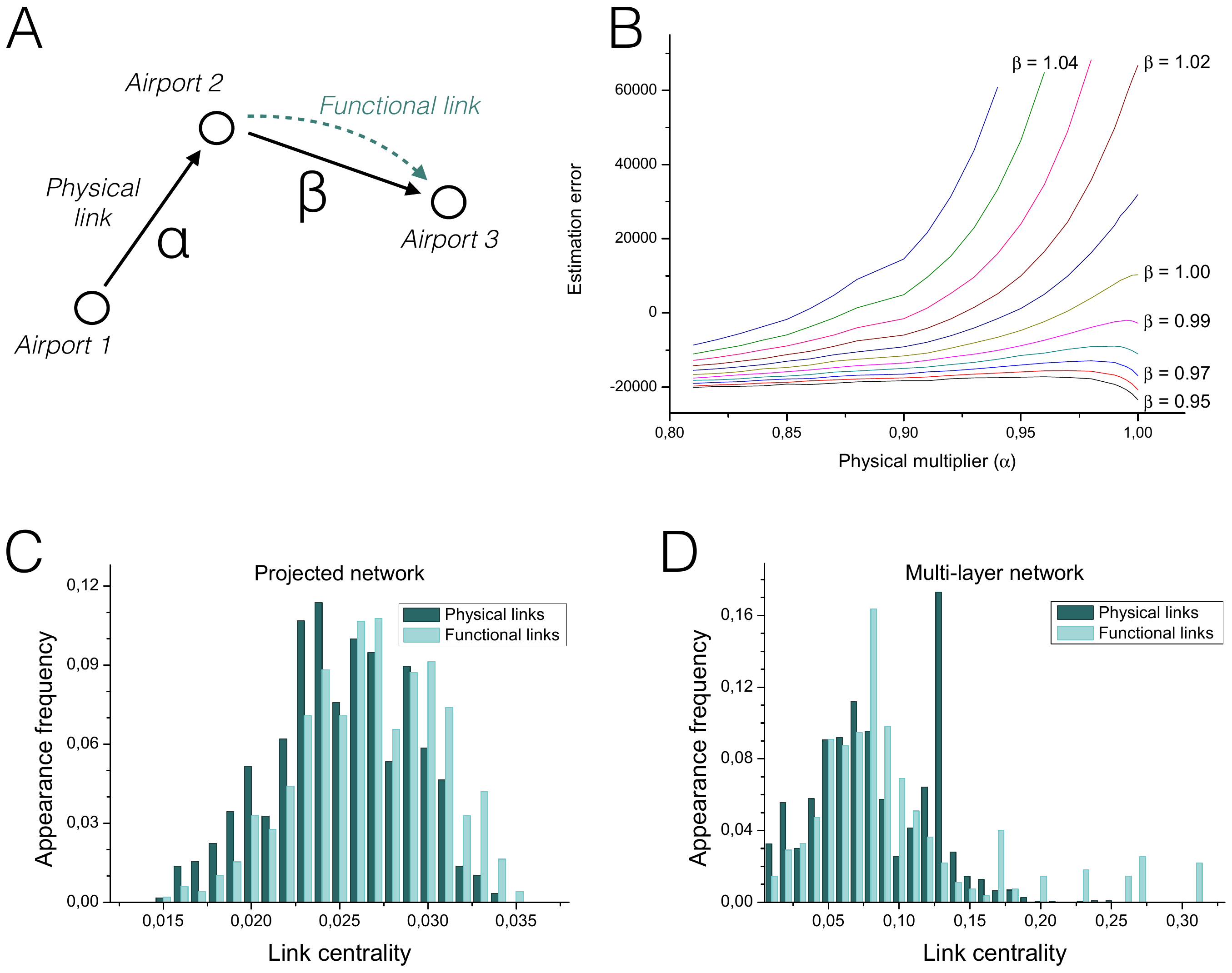}
\caption{Construction and results of the dynamical model. ({\it A}) Nodes (airports) are connected both by physical links (black solid arrows) when aircraft are allowed to move between two nodes, and by functional links (green dashed arrow) when a functional relationship is detected between the corresponding delay time series. When both links are present, the delay multiplier has a value of $\beta$, and $\alpha$ otherwise. ({\it B}) Difference between the delay obtained in the multi-layer and the single layer representations, as a function of $\alpha$ and $\beta$. The two input parameters, $\alpha$ and $\beta$, are respectively represented by the $X$ axis and by different curves. To obtain the estimation error, one thus has to fix the $X$ position ($\alpha$), intercept the desired curve ($\beta$), and read the result in the $Y$ axis. ({\it C, D}) Probability distribution for links in the single and multi-layer networks; light and dark green respectively represent physical and functional links.}\label{fig:03}
\end{center}
\end{figure*}

In order to address these differences, a simple dynamical model has been constructed, mimicking the process of delay propagation in the real system. It comprises aircraft performing a random walk on top of the physical networks, with a stochastic delay added at each time step to simulate the appearance of uncorrelated disturbances. Aircraft delays are further modified according to a {\it delay multiplier} \cite{beatty1999}, which may take two different values: $\alpha$ when only a physical connection ({\it i.e.} a scheduled flight) is present between the start and destination nodes, and $\beta$ when a functional link is also detected - see \ref{app:model} and Fig. \ref{fig:03} A for further details. Thus $\alpha$ is associated to routes where the delay is expected to partly recover, while $\beta$ indicates some kind of delay propagation.
Two variants of the model are also considered. In the first one, all aircraft share the same network, {\it i.e.} the single-layer projection is used; in the second, the full multi-layer structure is implemented, allowing aircraft to respect the corresponding airline connection network.

Once defined this two-fold model, it is relevant to study the error introduced by using the single layer representation, in terms of the amount of delay accumulated during the operation. In other words, we are interested in estimating the quantity $D _{ML} - D _{SL}$ (see \ref{app:model} for the definition), and in describing how it evolves as a function of the two parameters $\alpha$ and $\beta$. The resulting estimation error is presented in Fig. \ref{fig:03} B as a function of $\alpha$ and $\beta$. For large values of $\beta$, the delay obtained in the multi-layer representation increases to twice the value obtained in the single-layer case. Such error is due to the different topology of both networks: specifically, a significant number of functional links ({\it i.e.} associated to the $\beta$ multiplier) in the multi-layer representation have a high centrality, thus indicating that key airline connections are also the ones responsible for the delay propagation; on the other hand, such connections may loose importance in the projected networks, thus reducing the amount of delay generated
(see Fig. \ref{fig:03} C and D).

\begin{figure*}
\begin{center}
\includegraphics[width=0.99\textwidth]{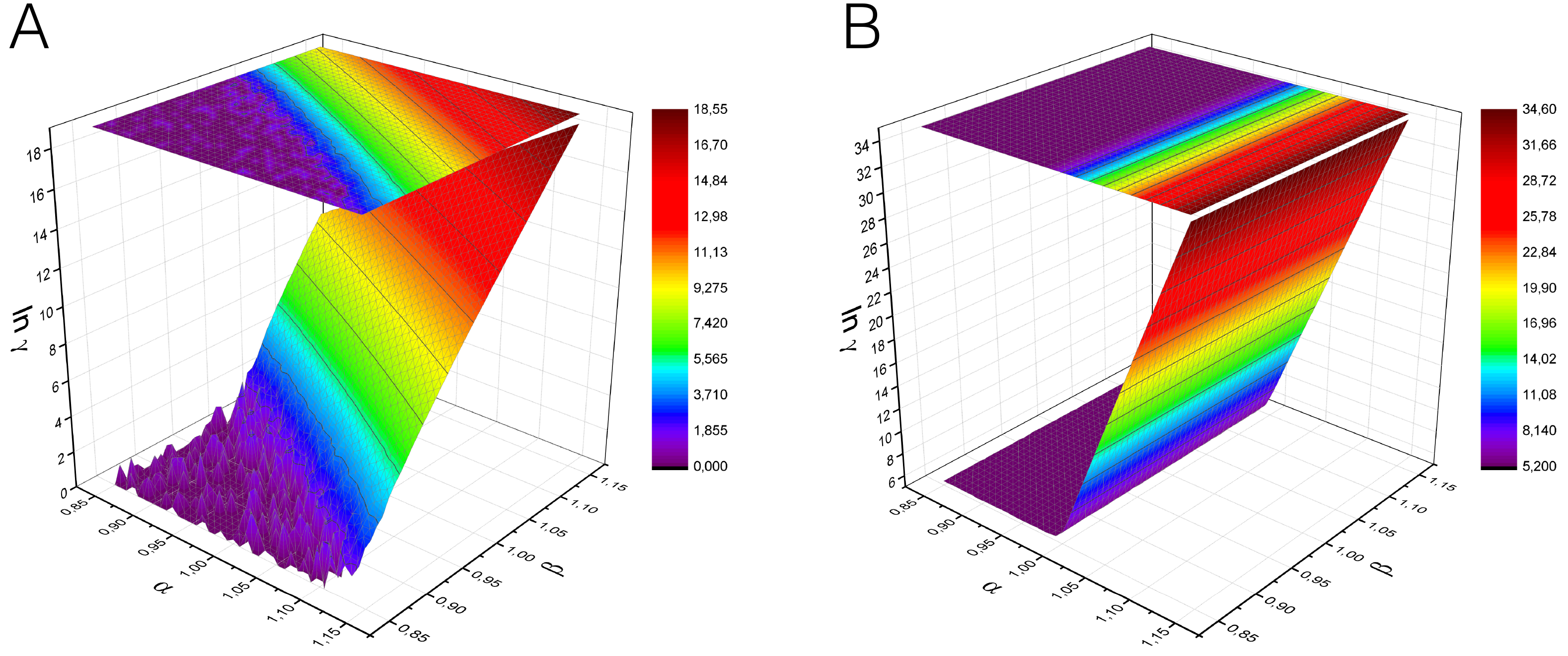}
\caption{Asymptotic dynamics of the dynamical model. Panels {\it A} and {\it B} respectively represent $\ln \gamma$ as a function of $\alpha$ and $\beta$, for the single layer and multi-layer models. For the sake of clarity, values of $\gamma$ smaller than $1$ have been set to $1$, such that $\ln \gamma \geq 0$.}\label{fig:04}
\end{center}
\end{figure*}

In order to better understand the differences between the single and multi-layer scenarios, the average increment of delays (denoted by $\gamma$) has been calculated - see \ref{app:model}. Fig. \ref{fig:04} reports the value of $\ln \gamma$ as a function of $\alpha$ and $\beta$ - the two delay multipliers, as in Fig. \ref{fig:03}. The single and multi-layer scenarios (respectively left and right panels) present two distinct behaviors. Specifically, the change from a steady state ($\gamma \approx 0$) to a regime of increasing delays ($\gamma > 0$) depends both on $\alpha$ and $\beta$ when the dynamical model is executed on a single-layer projection. On the contrary, the behavior of the multi-layer model strongly depends on the value of $\alpha$, $\beta$ having a neglectable impact.

\section{Discussion}

While functional networks have represented a great leap forward in the understanding of complex systems, the most notable of them being the human brain, little attention has been devoted to the effects of collapsing multi-layer dynamics into a single layer functional graph. When the dynamics of different layers have similar characteristics, the use of a projection is expected to have negligible effects on the results; nevertheless, when layers are heterogeneous in nature, such simplification may be misleading.

In this paper, I tackle this issue by analyzing the problem of delay propagation in the European air transport network. Different functional network representations are compared, {\it i.e.} the complete multi-layer representation, and the corresponding projections of the topology and dynamics. When analyzing their topological characteristics, important differences are observed. Notably, the most central nodes in the projections do not correspond to the nodes of high centrality in each layer; therefore, the former analysis give biased estimations, which cannot reliably be used to detect the most critical elements in the system. Furthermore, when a simple dynamical model is executed, the magnitude of the error yielded by considering a single layer projection is as big as the results themselves, thus indicating that any estimate obtained with this simplification is meaningless.

These results present important implications for the modeling and forecasting of the air transport network. First, they imply that any simulation performed to understand the dynamics of the system may yield misleading results when the multi-layer structure created by airlines is neglected. In spite of this, most of the recent research works in this fields fail to include this essential ingredient, both in the analysis of delay propagation \cite{beatty1999, sridhar2008modeling} and of the network robustness to disruption and attacks \cite{lacasa2009jamming, lordan2014robustness, lordan2015robustness}.
Second, it has to be noticed that the air transport system is created by the interactions between a large number of agents, which may create different layers along different {\it dimensions}. For instance, multiple flights do not just share the airline, but they may also be connected by the crew operating them. Disregarding these different {\it layer dimensions}, {\it e.g.} crews, aircraft types or flight type (cargo or passengers), may further bias our understanding of the system. While it may not be possible to include all these elements in one analysis, for instance due to the lack of publicly available data, it is safe to conclude that ``one layer does not fit all''.

Beyond the air transport network, the results here reported can be applied to any other multi-layer complex system analyzed by means of functional networks: the most paradigmatic example is then the study of the brain.
While the air transport network and the human brain are different in nature, some results obtained for the former can be translated into the analysis of the latter.
Specifically, both systems are characterized by a {\it transport layer} in which items are transported between nodes. The structure created by direct flights is conceptually similar to the {\it connectome}, {\it i.e.} the structure created by fibers connecting different regions of the brain, which allows information (spike trains) to move between them \cite{hagmann2010}.
Both systems can also be analyzed from an {\it information processing} point of view: the air transport processes information about delays, as transported by flights, while the brain information about the body and the environment, codified by neural spikes.
If the most important airports, in term of delay propagation, cannot reliably be detected with a projected functional network, the identification of functional hubs in the brain dynamics may be confused by the fact that the multi-layer structure of the cortex is neglected. Therefore, global hubs may not correspond to the most important nodes in each layer: the single layer analysis may then be misinforming about the real structure created by information flows.
In order to confirm and complete our knowledge about brain dynamics, it is then desirable to develop techniques able to accurately represent its multi-layer structure and dynamics.

\section{Acknowledgements}
This project has received funding from the European Union’s Seventh Framework Programme for research, technological development and demonstration under grant agreement no. 314087.

\appendix

\section{Air traffic data}
\label{app:data}

Information about aircraft trajectories has been extracted from the Flight Trajectory (ALL-FT+) data set as provided by the EUROCONTROL PRISME group. It includes information about planned and executed trajectories for all flights crossing the European airspace, with an average resolution of 2 minutes. The data set covers the period from 1$^{st}$ March to the 31$^{st}$ December 2011, including a total of 10.3 million flights. From this, a subset has been considered, composed of the 50 busiest airports and the 20 largest airlines (in number of flights) operating at these airports.

\section{Time series extraction}
\label{app:TSExtraction}

For each combination of airport / airline, a time series has been extracted, representing the average hourly delay of arriving flights. Due to some missing days, each time series comprises $7440$ values. The resulting time series presented several trends, as strong delays are expected mainly at peak hours, during week days, and during summer, {\it i.e.} those periods in which the traffic is higher. A detrend process has then been performed, by subtracting the average delay observed in the same day, in the two previous and posterior weeks, at the same hour, {\it i.e.}:

\begin{equation}
\bar d(t) = d(t) - \frac{1}{4} \sum\limits_{i \in \{  - 2, - 1,1, - 2\} } {d(t + 168i)}, 
\end{equation}

$d(t)$ being the original time series at time $t$, and $\bar d(t)$ the final time series. According to this definition, $\bar d(t)$ thus represents the difference between the observed and the expected (historical) delay.

Functional networks have been reconstructed by considering all pairs of airports, and by assessing the presence of a relationship among their time series by means of four metrics:

\begin{description}

\item[Pearson's linear correlation (LC):] a well-known measure of linear correlation, is defined as the covariance of the two variables divided by the product of their standard deviations \cite{pearson1895}.

\item[Mutual Information (MI):] measures the information shared by two random variables, $X$ and $Y$, by analyzing their individual and joint probability distribution functions \cite{li1990}:

\begin{equation}
I(X;Y) = \sum\limits_{y \in Y} {\sum\limits_{x \in X} {p(x,y)\log \left( {\frac{{p(x,y)}}{{p(x)p(y)}}} \right)} } 
\end{equation}

\item[Granger Causality (GC):] measures the presence of a causal relation, {\it i.e.} if one time series is useful in forecasting a second one \cite{granger1969}. Given two time series $X$ and $Y$, $X$ {\it causes} $Y$ if $X$ values provide statistically significant information about future values of $Y$. This is usually calculated by means of univariate autoregressions, and tested through t-tests and F-tests.

\item[Transfer Entropy (TE):] given two processes (or time series) $X$ and $Y$, TE is the amount of uncertainty reduced in future values of $Y$ by knowing the past values of $X$ and $Y$: 

\begin{equation}
\begin{split}
T_{X \to Y}  = H\left( {Y^t \left| {Y^{t - 1:t - L} } \right.} \right) - \\
H\left( {Y^t \left| {Y^{t - 1:t - L} } \right.,X^{t - 1:t - L} } \right),
\end{split}
\end{equation}

$H$ being the Shannon's entropy of the time series.
While similar in nature, the advantage of TE with respect to GC is that the former can be used in the analysis of non-linear signals \cite{schreiber2000}.

\end{description}

These four synchronization metrics have been chosen to cover different characteristics of the time series: they include two correlations (LC and MI) and two causalities (GC and TE); furthermore, two of them are linear (LC and GC), two non-linear (MI and TE).
Results obtained with these four metrics are qualitatively similar, thus only the ones corresponding to TE are here reported.

Finally, a physical network has been reconstructed, where nodes represent airports, pairs of them being connected by unweighted links when at least one direct flight has been operated between them. Two variations of this network are considered for constructing the dynamical model (see next Section): the complete multi-layer structure, each layer representing the network of flights of a single airline, and a single layer projected network.

An interactive representation of the obtained networks, both physical and functional, can be found at \cite{Interactive}.

\section{Dynamical model}
\label{app:model}

In the dynamical model representing delay propagation through the air transportation system, a set of $n = 10.000$ aircraft are randomly placed in the $50$ nodes of the network; in the case of the multi-layer representation, aircraft are also randomly assigned to one of the 20 layers available. Initially, all aircraft are on time, {\it i.e.} their delays $d_i (t=0) = 0$ with $i \in [1, n]$. At each time step $t > 0$, aircraft are randomly moved according to the physical adjacency matrix, corresponding to the flight connections detected in the system, respecting the assigned layer in the case of the multi-layer structure; furthermore, their delays evolve according to:

\begin{eqnarray}
d_i (t) = \alpha d_i (t-1) + \epsilon ,  \text{if } a = 0,\\
\label{eq:modelA}
d_i (t) = \beta d_i (t-1) + \epsilon ,  \text{if } a = 1,\\
\label{eq:modelB}
d_i (t) = 0, \text{if } d_i (t) < 0,
\label{eq:model}
\end{eqnarray}

$\epsilon$ being a random number drawn from a normal distribution $\mathcal{N}(0, 1)$, and $a$ representing the presence of a functional link between the start and destination nodes (see Fig. \ref{fig:03} A). The last condition simulates the scheduling of flights: when an aircraft reaches its destination airport before the scheduled time, it waits until the expected time of departure.

The two parameters $\alpha$ and $\beta$ define how delays are expected to evolve after every flight, and are commonly known in the aeronautical community as {\it delay multipliers} \cite{beatty1999}. A value larger (smaller) than $1.0$ indicates that each flight is, on average, arriving with a delay that is larger (respectively, smaller) than the one recorded at take-off, and is thus multiplying (dampening) the total delay. In order to simulate this process in the dynamical model, each possible route is associated to either $\alpha$ or $\beta$, depending on the existence of the corresponding functional link between the departure and destination airports (see Eq. \ref{eq:modelA} and \ref{eq:modelB}). When $a = 0$, {\it i.e.} no causality link has been detected between two airports, the delay propagation is controlled by $\alpha$; the lack of a causality relation suggests that delays are (at least partly) recovered, and thus $\alpha$ is expected to be smaller than $1.0$. On the other hand, if $a = 1$, a delay propagation process is probably present between the two airports: $\beta$ is thus expected to be greater than $1.0$, and also that $\beta > \alpha$.
It should be noticed that the single and multi-layer networks may have a different number of physical and functional links. In order to have comparable results, in the latter case the model is executed with two parameters $\alpha^*$ and $\beta^*$ defined as:

\begin{eqnarray}
\alpha^* = \alpha ^ { p ^{ML} / p ^{SL} }, \\
\beta ^* = \beta  ^ { f ^{ML} / f ^{SL} },
\end{eqnarray}

$p$ and $f$ respectively being the proportion of physical and functional links in the multi-layer ($ML$) and single layer networks ($SL$).

The final delay generated by the system is measure as:
\begin{equation}
D = \sum _{i} d_i( t = 100 ).
\end{equation}

Results reported in Fig. \ref{fig:03} B correspond to the average total delay $D$ obtained in $100$ realizations of the model.

The importance of each link in the networks (Fig. \ref{fig:03} C and D) is calculated as the $\alpha$-centrality of the corresponding dual graph, where link weights are normalized such that the out-strength of each node is equal to one - thus simulating a random walk on the network.

Finally, the average increment of delays has been calculated as:

\begin{equation}
\gamma = \langle \sum\limits_{i} d_i(t) - \sum\limits_{i} d_i(t - 1) \rangle,
\label{eq:gamma}
\end{equation}

$d_i(t)$ being the delay observed at time $t$ for airport $i$, and $\langle \cdot \rangle$ the average through time. Values of $\gamma$ close to $0$ imply that the system is able to dissipate the delays generated by the random part of Eqs. 4-6; on the contrary, for $\gamma > 0$, a steady state is not reached and delays constantly increase.

\bibliographystyle{model1-num-names}
\bibliography{Multilayer}{}

\end{document}